\documentclass[%
 reprint,
 amsmath,amssymb,
 aps,
floatfix
]{revtex4-2}

\usepackage{graphicx}
\usepackage{dcolumn}
\usepackage{bm}

\usepackage{xcolor}

\newcommand{\blue}[1]{\textcolor{blue!80!black}{#1}}

\DeclareMathOperator*{\argmin}{arg\,min}

\usepackage{svg}
\usepackage{amsthm}

\usepackage{amsmath}
\begin{document}

\title{An MCMC method to determine properties of Complex Network ensembles}
\title{\blue{Short but meaningful title containing "Canonical Network Ensembles"}}
\title{\blue{Canonical Network Ensembles and their application}}
\title{\blue{Canonical Network Ensembles approach to Small-World networks}}
\title{Relative Canonical Network Ensembles -- \\(Mis)characterizing Small-World Networks}

\author{Oskar Pfeffer, Nora Molkenthin, Frank Hellmann}
 \affiliation{Potsdam Institute for Climate Impact Research\\ TU Berlin}
\date{\today}

\begin{abstract}

What do generic networks that have certain properties look like? We define Relative Canonical Network ensembles as the ensembles that realize a property R while being as indistinguishable as possible from a generic network ensemble. This allows us to study the most generic features of the networks giving rise to the property under investigation. To test the approach we apply it first to the network measure "small-world-ness", thought to characterize small-world networks. We find several phase transitions as we go to less and less generic networks in which cliques and hubs emerge. Such features are not shared by typical small-world networks, showing that high "small-world-ness" does not characterize small-world networks as they are commonly understood. On the other hand we see that for embedded networks, the average shortest path length and total Euclidean link length are better at characterizing small-world networks, with hubs that emerge as a defining feature at low genericity. We expect the overall approach to have wide applicability for understanding network properties of real world interest.
\end{abstract}

\maketitle

\section{Introduction}
Network ensembles are sets of networks together with a probability distribution of their occurrence and have been successfully used to model a wide range of natural, social and technical systems, in which the interaction structure is subject to, or the outcome of, stochasticity \cite{schultz2014random,opitz1996generating,snoeijer2004force,dorogovtsev2000structure,molkenthin2016scaling,molkenthin2018adhesion}. Typically those ensembles are generated through a heuristic process, thought to capture some aspect of the microscopic formation process, which underlies the real-world system they are trying to model. The resulting ensemble can then be studied and characterized by means of network measures that quantify certain properties of the networks. Examples for this are Watts--Strogatz networks, which are characterized by low average shortest path length and high clustering coefficients \cite{watts1998collective}, and Barabasi--Albert networks, which are characterized by their power-law degree distribution \cite{barabasi1999emergence}.

Here we want to approach network ensembles from the other side. Rather than trying to model real world networks we ask: What do generic networks that have certain properties look like? Thus, we will \emph{define} ensembles through a particular property captured by a ``property function'' $R(G)$ on networks $G$ and a background ensemble that defines our notion of generic networks in the given context. To this end, we will consider slightly generalized exponential random graphs. Exponential random graphs have long been a tool in network science, starting with \cite{holland1981exponential,frank1986markov,strauss1986general,newman2003structure}, see  \cite{cimini2019statistical} for a recent review, and are also sometimes known as canonical network ensembles (CNE)
\cite{park2004statistical,bianconi2007entropy,bianconi2009entropy}. We will consider CNEs relative to the background ensemble of generic networks. Given some set of networks $\mathcal{E}$ on a finite set of vertices, denote the probability distribution of the background ensemble as $q(G)$ for $G \in \mathcal{E}$. The relative canonical network ensemble (RCNE) of $R$ relative to $q$ is given by the probability distribution proportional to $\exp(- \beta R(G)) q(G)$.

We emphasize that our aim is not to model empirically observed network ensembles with certain properties. There is no reason to expect empirical networks, that are the outcome of subtle formation processes, to be generic. Instead, we will study the properties themselves, specifically the most generic features that produce them, and whether or not the properties suffice to generically characterize the networks under study. Our aim in this is to understand properties that are of considerable practical interest. Companion papers will consider epidemic thresholds and the vulnerability to failure cascades in power grids. To introduce our approach, this paper will focus on well-known and well-established network measures, that are computationally challenging, instead. Specifically, we will consider the notion of "small-world-ness".

We study two ensembles, the first defined by the \emph{small-world-ness}, as introduced in \cite{Humphries2008}, the second defined by a combination of Euclidean link length and average shortest path length similar to \cite{Mathias2001}. To study these ensembles we sample them using the straightforward Metropolis-Hastings (MH)\cite{Metropolis1953EquationOS, Hastings1970MonteCS, iba2014multicanonical} algorithm.

In both cases we find phase transitions as we go from fully generic networks to highly specific ones. At these phase transitions certain features arise, e.g. hubs and cliques start appearing in the ensemble. Surprisingly we find that generic networks with high small-world-ness do not resemble small-world networks. Thus, we find that what \cite{Humphries2008} called small-world-ness does not actually characterize small-world networks generically.

\section{Relative Canonical network ensemble}

Exponential random graphs were first introduced in \cite{strauss1986general,holland1981exponential,frank1986markov}. Given the set of simple graphs $\mathcal{E}_N$ on a set of $N$ vertices, they are defined by the probability distribution over $\mathcal{E}_N$, \( p_{\beta}^R(G) = {Z_R(\beta)}^{-1} \exp\left(-\beta R(G) \right) \). That is, they are the Gibbs ensemble at temperature \( T = 1/ k_B \beta \). The use of such network ensembles is sometimes justified by the fact that these are maximum entropy ensembles with a given expectation value for $R$. However, there is no a priori reason to expect formation processes that lead to real world networks to maximize entropy. For instance, typical formation processes do not resemble exchange with an environment at fixed genericity (in analogy to a heat bath). In fact, it was already noted in \cite{newman2003structure} that the maximum entropy ensembles do not model real world systems easily and show unexpected structures, interpreted there as an ``unfortunate pathology''.

Instead, we  want to understand the most generic features giving rise to a property $R$. That is, a feature, that is observed more frequently the more the expectation value of $R$ differs from the value expected for generic networks. As mentioned in the introduction, to define our notion of genericity we specify 
background ensemble $q(G)$ (for example an Erdős–Rényi ensemble at a fixed number of edges). The \textit{relative canonical network ensemble} of $R$ relative to $q$ is then given by:
    
\begin{align}
p^R_{\beta, q}(G) = \frac1{Z_{R,q}(\beta)} e^{-\beta R(G)} q(G)\; ,
\end{align}

with normalization/partition function $Z_{R,q}(\beta) = \sum_{G \in \mathcal E_N} e^{-\beta R(G)} q(G)$.
This ensemble is characterized by being the ensemble of minimum relative entropy $D(p || q)$ for a fixed expectation value of $R$. From an information-theoretic perspective it is the ensemble hardest to distinguish from the generic ensemble $q$ while having fixed expectation value $\langle R \rangle = R^*$, for a more detailed discussion see Appendix \ref{Sec::Relative Entropy}.

\begin{figure}
    \centering
    \includegraphics[width=\columnwidth]{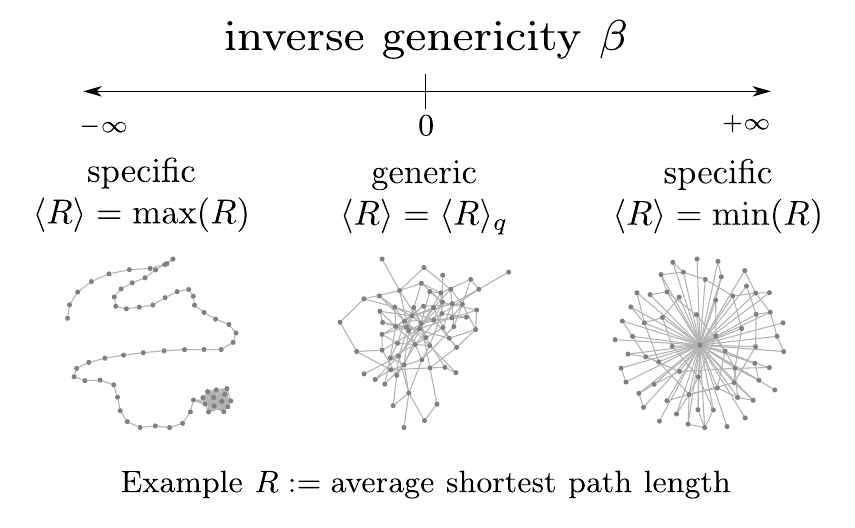}
    \caption{The inverse genericity $\beta$ mediates between specific ensembles concentrated on maximum $R$, the generic background ensemble $q$ with $\langle R \rangle = \langle R \rangle_{q}$ and specific ensembles concentrated on minimum $R$.}
    \label{fig:genericity}
\end{figure}

The parameter $\beta$ moderates the trade-off between the generic ensemble and highly specific ones peaked on networks that are high or low in $R$, see Figure~\ref{fig:genericity}. It can range from $-\infty$ to $+\infty$ with the sign depending on whether the expectation value of $R$ is higher or lower than in the generic network ensemble given by $\beta = 0$. At $\beta \rightarrow -\infty$  we have an ensemble concentrated on $\max(R)$, while at $\beta \rightarrow +\infty$ it is $\min(R)$. This, and the fact that interpretation of the relative entropy is purely information theoretic (rather than thermodynamic), motivates us to refer to $\beta^{-1}$ in this context as the genericity rather than as a temperature.

Of particular note are phase transitions that occur as we lower the absolute genericity. 
The structure of the ensemble changes at and beyond the phase transition. This change in structure allows us to identify specific features that contribute to property $R$ but are not generic enough to occur before.

Throughout the rest of this manuscript we will 
consider canonical ensembles relative to the Erdős–Rényi ensemble at fixed size $N$ and mean degree $k$, that is, the equidistribution over all graphs with vertex set $\{1,...,N\}$ and $kN/2$ edges. Generally what counts as a generic network is highly dependent on context. A generic social network does not look like a generic power grid. In some contexts it might also be appropriate to use maximum entropy null-models as generic ensembles\cite{cimini2019statistical}.

Since exponential random graphs were first introduced, computing capabilities profoundly increased. This means we can now use complex, practically relevant network properties and analyze what features of networks generically give rise to them. This approach may help in the future to gain a better understanding of complex network measures and provide a way to find simpler network measures to act as predictors for the characteristics defining the ensemble.

To study these ensembles we need to sample from them. An important property of RCNEs is that they are well suited for sampling using Metropolis-Hastings (MH) algorithms. To use MH on our relative ensemble, we require a background process that generates proposed steps compatible with the background distribution $q$. For $q_{Nk}$ this can be provided simply by considering rewiring of edges. Starting from a system in state \( x \) the algorithm proposes rewirings that are accepted with probability 
\begin{align}
    P_\beta(x \rightarrow y) = \min \left( 1, \frac{p_{\beta}(R(x))}{p_{\beta}(R(y))} \right) = \min \left(1, e^{-\beta \Delta R}\right).
    \label{eq:kernel}
\end{align}

This algorithm satisfies the detailed balance condition and the Markov chain defined by it is strongly connected. In the limit of infinite steps the time average for this ensemble converges to the ensemble average of the ground state which is the relative canonical network ensemble. Unfortunately, there are no guarantees for finite time samples and we have to resort to heuristics to understand whether convergence has occurred. To do so we typically also run several chains in parallel from random initial conditions. This further allows us to obtain less correlated samples. More details on our sampling approach are provided in the next section.

For more general background ensembles it might be complicated to find step proposals. If $q$ is explicitly known it can be incorporated into the step acceptance probability. If there is no explicit formula for $q$, for example because it is implicitly defined by a stochastic growth process, it is necessary to make use of the growth process directly to generate new proposals.

\section{Small world properties}
\begin{figure}[t!]
    \centering
    \includegraphics[width=\linewidth]{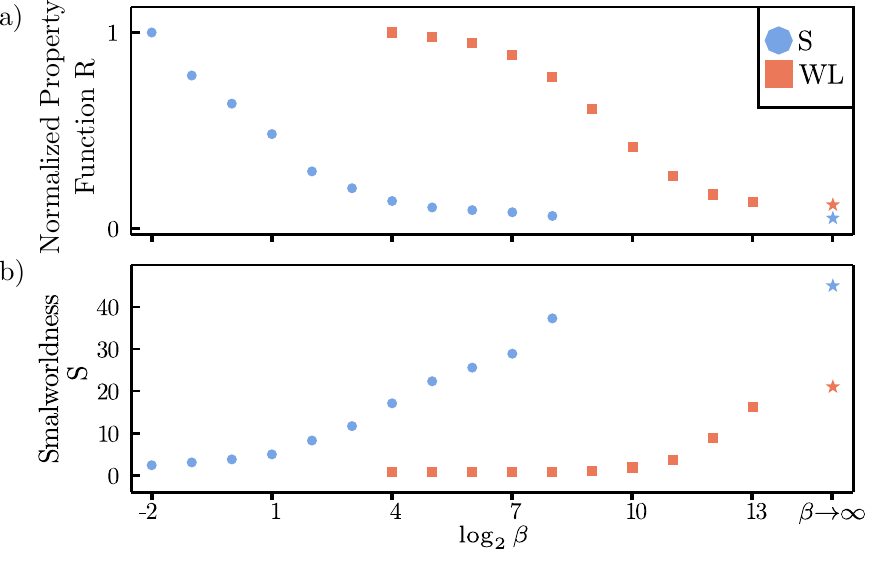}
    \caption{\textbf{Small-world-ness increases significantly as the property function approaches the global minimum.} $R_S$-ensemble (circles) and $R_{WL}$-ensemble (squares) networks with \(N=256\) and \( \langle k \rangle=4\) to finite inverse genericity \(\beta \in \{2^{-2} \to 2^{13}\}\). At the start of the range the ensembles are statistically indistinguishable from the background ensemble $q$. a) shows the small-world-ness \(S\) and b) the property \(R_{S}\) and \(R_{WL}\) vs \(\beta\). The stars on the right identify simulations for \(\beta\rightarrow\infty\). Each data point is averaged over 32 realizations with \(2^{24}\) MCMC steps each.
    }
    \label{fig:Q_R_SWN}
\end{figure}

To demonstrate the approach, we analyze two different small-world-ness properties. In particular we look at the features that give rise to them and whether they generically characterize what is commonly known as small-world networks. In the first instance we consider the Small-world-ness measure \(S = ({C}/{L}) \left( {C_{\text{ER}}}/{L_\text{ER}} \right) ^{-1}\), introduced in~\cite{Humphries2008}, where \( C \) is the global clustering coefficient defined as the number of closed triplets divided by the number of all triplets, \( L\) the average shortest path length $C_{\text{ER}}$ is the average clustering coefficient of an Erdős–Rényi network \cite{erdds1959random} of the same size and $L_{\text{ER}}$ is the expected average shortest path length of an Erdős–Rényi network of the same size.
Finally, the form of the property we will consider is:

\begin{align}
    R_{S} = \frac{L}{C}.
\end{align}

That is, proportional to the inverse of $S$. Thus, small values of $R_{S}$ indicate high Small-world-ness, and we are interested in positive $\beta$.
\begin{figure}
    \centering
    \includegraphics[width=\columnwidth]{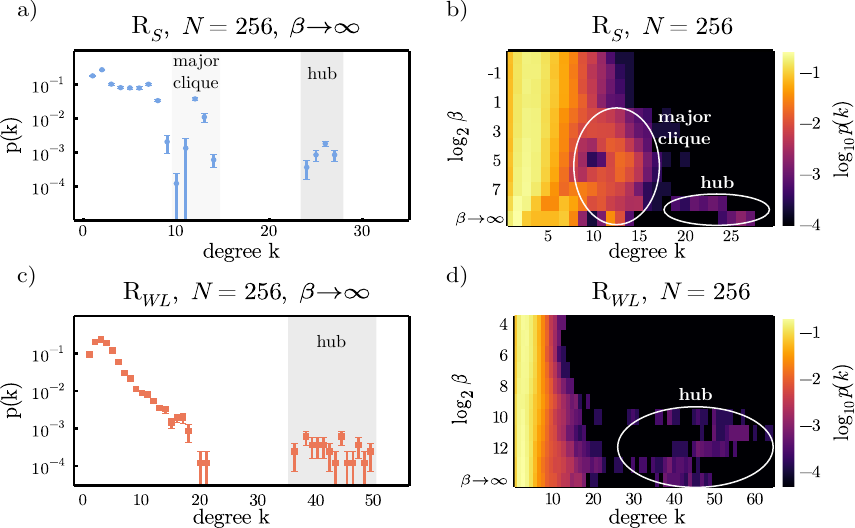}
    \caption{\textbf{At low genericity hub and clique structures emerge, transforming the degree distribution.} The degree distribution in a) shows three peaks for the $R_{S}$-ensemble while the degree distribution of the $R_{WL}$-ensemble in c) shows two peaks. b) and d) show a heat-map of degree distributions at various inverse genericities. The degree distributions are an average of 32 realizations with \(2^{24}\) MCMC steps each.}
    \label{fig:degdist}
\end{figure}

The second property

\begin{align}
    R_{WL} = W L
\end{align}

is given in terms of the average shortest path length \( L \) and the wiring length \(W\) in an embedded network. \(W\) is given by the sum of the Euclidean length of all edges. The networks for this ensemble are embedded in a 2D plane. The introduction of \(W\) was inspired by \cite{Mathias2001}, where it was argued that small-world networks might arise as a secondary feature from a trade-off between maximal connectivity and minimal edge lengths. Again small $R_{WL}$ is expected to yield Small-world networks and we consider positive $\beta$.

\begin{figure*}[t!]
    \centering
    \includegraphics[width=\textwidth]{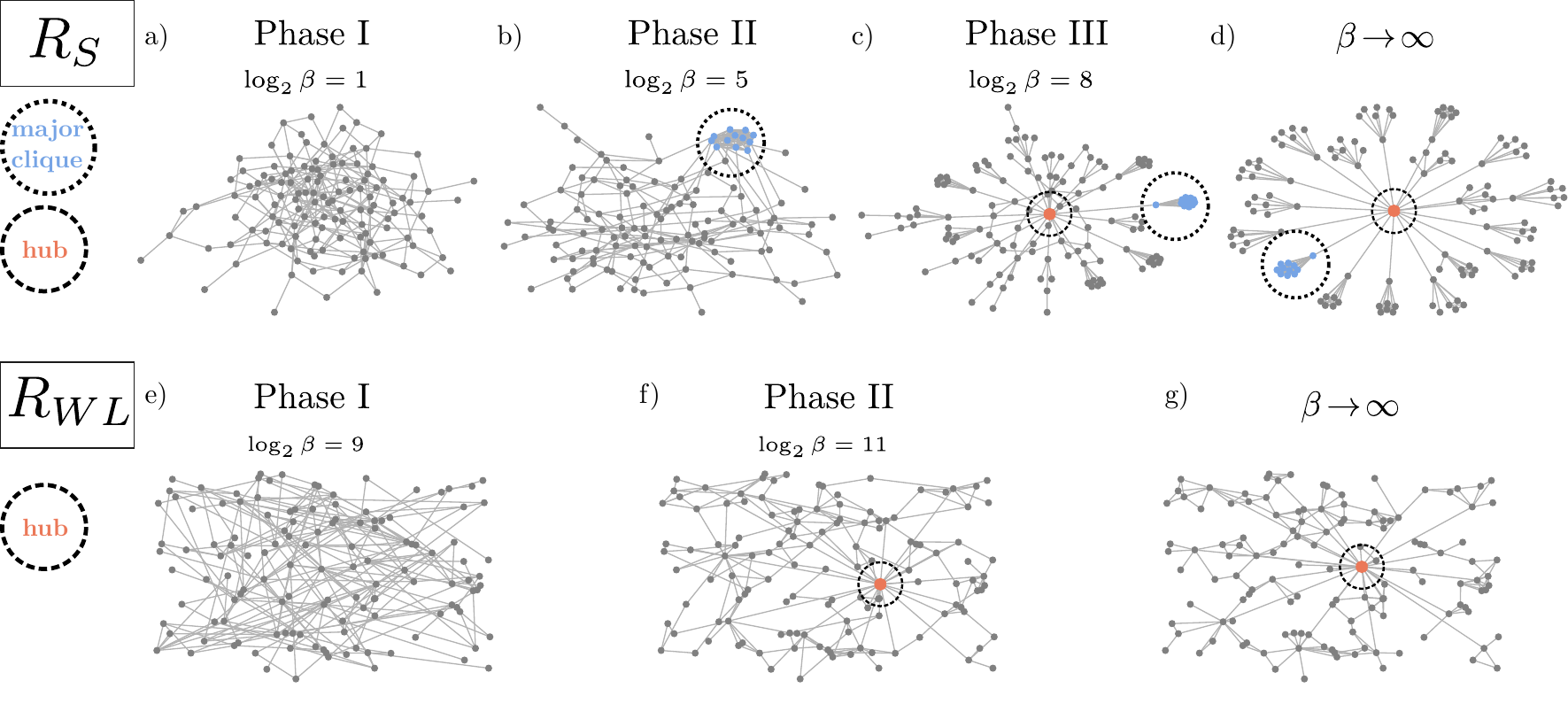}
    \caption{\textbf{Examples of the various phases for $N=128$ networks.} The $R_S$ ensemble starts as a random network without any recognizable structure in a), then a first large clique emerges in b) and finally a central hub that lowers the average shortest path length in c), which is close to the network where the small-world-ness is maximal in d). The $R_{WL}$ ensemble starts from a random phase in e), then a central hub with long range connections emerges in f). The network minimizing $R_{WL}$ resembles a random geometric network with a central hub.}
    \label{fig:phase_examples}
\end{figure*}
Both ensembles are taken relative to a random Erdős–Rényi network with \(N\) vertices and \(M = {\langle k \rangle N}/{ 2} \) edges, where \(\langle k \rangle \) is the average degree of the network. The positions of the vertices in the embedded networks are initialized randomly on a 2D unit square.

The proposal for each Monte Carlo step is generated by rewiring a single edge, i.e. deleting an existing edge at random and connecting two previously unconnected vertices chosen at random, thus keeping the number of edges constant. The proposal is then accepted with the transition probability given in Eq.~\eqref{eq:kernel}. Proposals of disconnected graphs are always rejected since \(L\) is infinite. To ensure convergence at low genericity, we use an exponential schedule \(\beta^{-1} (t) = \left( {\beta^{-1}_\text{start} \alpha ^t + \beta^{-1}_\text{end}} \right) \), similiar to the Simulated Annealing approach \cite{Kirkpatrick}, where \(t\) is the step parameter, \( \alpha =0.99 \) is a simulation parameter.
\(\beta^{-1}_\text{start} \) and \(\beta^{-1}_\text{end}\) are the start and final genericities. 
We generated ensembles with $(128, 128, 128, 128, 64, 32, 16)$ networks of size $N=(8, 16, 32, 64, 128, 256, 512)$ correspondingly with average degree \(\langle k \rangle=4\). All the simulations appeared to converge, allowing qualitative evaluation of the simulations. 
Throughout the manuscript, genericity was decreased over $2^{11}$ equally long periods, each containing $2^{13}$ MCMC steps for a total of \(2^{24}\) MCMC steps. Note that while the properties we consider here are conceptually simple, the presence of the average shortest path length, which needs to be recomputed for every proposed step, renders them computationally expensive. We further note that achieving convergence for the $R_S$ ensemble was considerably harder than for $R_{WL}$. Thus, they do constitute a real check of the ability of the approach to study complex network properties.

As shown in Fig.~\ref{fig:Q_R_SWN}, the Small-world-ness \(S\) increases for both network ensembles as they become less generic. For the $R_S$-ensemble, for which it is mathematically guaranteed that the expectation value of $S$ increases for decreasing genericity, this is an important sanity check on our sampling. In the $R_{WL}$-ensemble this arises as a secondary effect as Euclidean and network distances are reduced, showing that generic $R_{WL}$ networks do indeed have high small-world-ness, as anticipated in \cite{Mathias2001}.

As a common and simple network measure, we now look at the degree distribution.
Fig.~\ref{fig:degdist} shows the degree distributions of the two ensembles for decreasing genericity. The shift from generic (poisson distributed) to specific networks is evident. The extremal $\beta \rightarrow \infty$ case (simulated with the same exponential schedule as above until we observed convergence) is shown explicitly in Fig.~\ref{fig:degdist} a) and c), and we see highly pronounced features in the degree distribution. Example of networks at this state are shown in Fig. \ref{fig:phase_examples} d) and g), looking at these allows us to identify the features in the degree distribution as major cliques and hubs. Note that the degree distribution of the $R_S$ ensemble in particular does not resemble that of the WS-ensemble.

The $R_S$ example network (Fig. \ref{fig:phase_examples} d) looks almost star-shaped with a very highly connected central node and a few fully connected branches. This indicates that the two components of the property, namely average shortest path length and clustering are optimized in specialized areas of the network. The star graph, which is the smallest possible sparse graph, is thereby combined with many nodes in fully connected cliques.
The $R_{WL}$ network (Fig. \ref{fig:phase_examples} g) on the other hand looks like a sparse geometric network with star-shaped shortcuts, making it much closer in spirit to the WS-ensemble and its two-dimensional relatives.

As a result, the nodes in the $R_S$-networks can be categorized into hub-nodes, clique-nodes, and the rest, where hub-nodes have very high degrees ($k \approx 20 - 30$), clique nodes above-average degrees ($k \approx 10 - 15$) and the rest has low degrees. This can be seen in Fig.~\ref{fig:degdist}b) as 3 major peaks.

To understand how these extremal cases come about, we consider the degree distributions over the whole parameter space in (Fig.~\ref{fig:degdist} b and d). Here we see several abrupt transitions. For the $R_{S}$ ensemble the major clique starts forming at $\beta\approx 2^{-2}$, while the hub only emerges at high inverse genericities of around $\beta\approx 2^{8}$. 

In the embedded networks, nodes fall into two categories: regional nodes and inter-regional hubs.
This can be seen in Fig.~\ref{fig:degdist} c), where regional nodes fall into the normal degree distribution of a (slightly sparser) geometric network and hubs have higher degrees of $k \approx 40 - 50$. This hub emerges at around $\beta\approx 2^{10}$.

\begin{figure*}[hbtp]
    \centering
    \includegraphics[width=\textwidth]{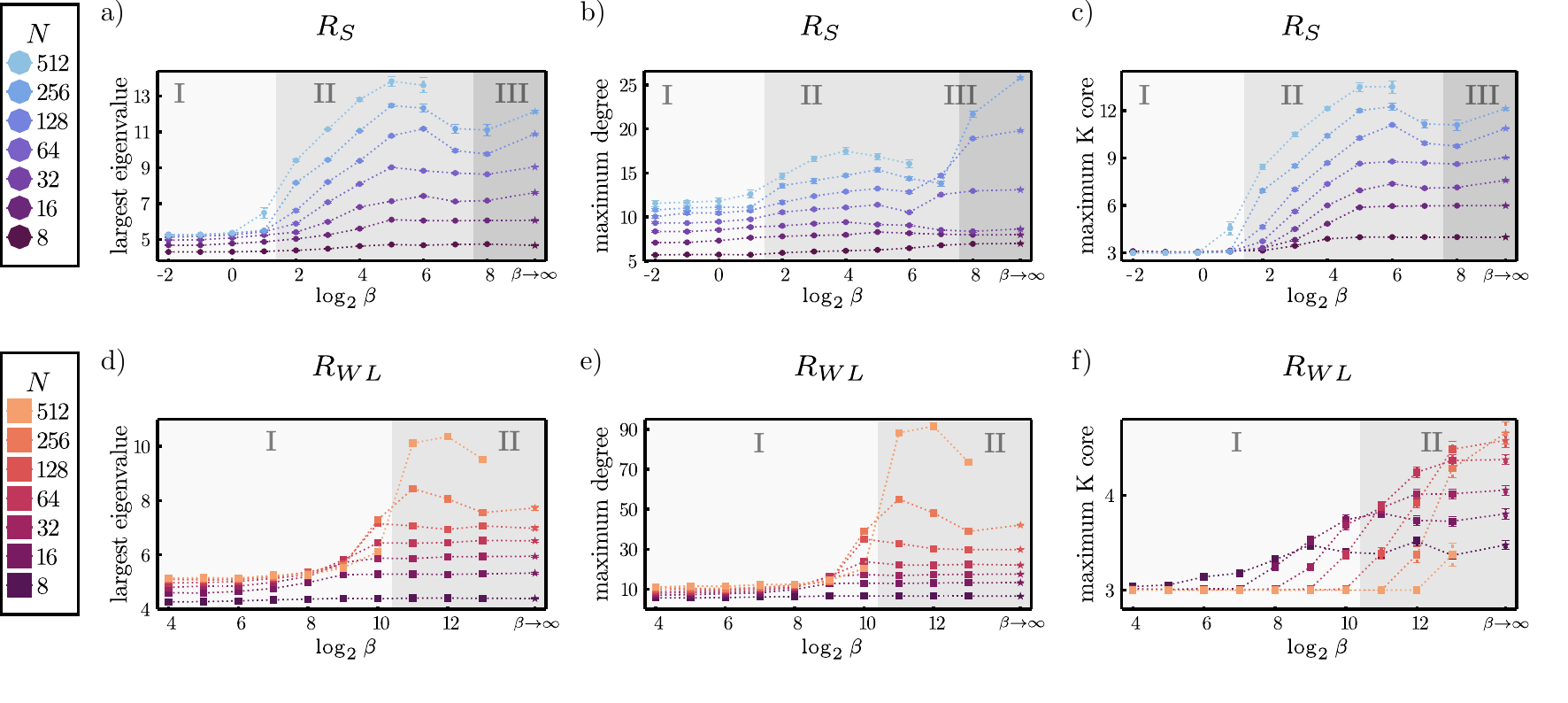}
    \caption{\textbf{The phase transition is characterized by a rise in the largest Eigenvalue.} The largest Eigenvalue is plotted in a) and d) for $R_{S}$ and $R_{WL}$ respectively. b) and e) show the dependence of the largest degree on the genericity and c) and f) show the maximum k-core over genericity for network sizes from $N=\{8, 16, 32, 64, 128, 256, 512\}$ and average degree $\langle k \rangle=4$. Simulation details: MCMC steps $=2^{24}$ each, generated ensemble size (in order of network size) $=\{128,128,128,128,64,32,16\}$. The realizations for $R_S,N=512, \log_2\beta\ge7$ did not fully converge and are not plotted.}
    \label{fig:transition}
\end{figure*}

\section{Genericity phase transition}

Fig.~\ref{fig:phase_examples} shows various examples of networks taken from different genericity phases. We can now study the transition between these phases in more detail.

As seen in Fig.\ref{fig:degdist} b and d, both the $R_S$ and $R_{WL}$ ensembles show a qualitative change in the degree distribution. At high genericity we have essentially random graphs in both cases with the expected Poisson degree distribution. At low genericity both ensembles show multiple peaks. In case of the $R_{WL}$-ensemble this comes as a sudden appearance of a second peak at $\beta=2^{10}$. In case of the $R_S$-ensemble this transition appears to be less clear cut with a structure that resembles branching at $\beta\approx 2^{2}$ and almost merging again, while another peak appears at $\beta\approx 2^{8}$.

To better understand these transitions we analyze the mean largest eigenvalues \(\lambda_1\) of the adjacency matrix, sizes of the largest non-empty $k$-core and maximum degree as functions of the genericity for network sizes from $N=2^3$ to $N=2^9$. The results are shown in Fig.~\ref{fig:transition}. 
Both ensembles show a phase transition in the largest Eigenvalue between a low $\lambda_1$ state and a high $\lambda_1$ state. This transition is located at $\beta \approx 2^{2}$ for the $R_S$-ensemble and at $\beta\approx 2^{10}$ for the $R_{WL}$-ensemble.

This transition is mirrored by the maximum k-core in case of the $R_S$-ensemble (see Fig.~\ref{fig:transition} c)). This indicates that here the formation of the first dense region in the graph is responsible for the phase transition. This is clearly not the case for the $R_{WL}$-ensemble, where we find no consistent transition genericity for the largest non-empty k-core, but a shift in its rise depending on the network size as shown in Fig.~\ref{fig:transition} f).

Instead, the largest Eigenvalue transition in the $R_{WL}$-ensemble is mirrored by the maximum degree in the network, as displayed in Fig.~\ref{fig:transition} e). As expected from the degree distributions shown above, the changes of the maximum degree in the $R_S$-ensemble hint at two transitions, one at $\beta\approx 2^{2}$, in which the first dense region forms and one at $\beta \approx 2^{8}$, at which the central hub forms. The phase transition giving birth to the first dense region found at $\beta\approx 2^2$, can be interpreted as similar to \cite{Park2005SolutionFT}, where a first order phase transition was analytically found for Strauss's model of clustering \cite{strauss1986general}.

These results show that certain discrete features suddenly emerge  at  certain  genericities. The transitions become qualitatively visible in the degree distribution, clearly appear in their graphical representations and can be quantified in various network measures, where the largest eigenvalue is a good first indicator and are more detailed in the maximum k-core and degree. These phase transitions and the emergence of hubs and cliques are a driving element in the increase of the small-world-ness property.

\section{Discussion and conclusion}
Here we introduced the concept of relative canonical network ensembles of arbitrary network properties, as a means to study what the most generic networks with these properties look like.
These ensembles are amenable to Metropolis-Hastings and MCMC methods, providing a simple and straightforward (if potentially computationally expensive) way of sampling from non-trivial network ensembles defined through network measures of practical interest.

To challenge the method we studied two properties traditionally expected to characterize small-world networks. Surprisingly we found that generic networks with a high small-world-ness index $S$ in the sense of \cite{Humphries2008}. Instead we find that as $S$ increases the most generic networks with high $S$ contain first cliques and then hubs, neither of which occur in the WS-ensemble. An alternative property defined as the product of wiring and shortest path length fared better, here also hubs arise for the least generic networks, but the system appears to resemble small world networks more closely. This indicates that at least for some networks, spatial embedding may actually be the defining feature, from which high small-world-ness arises as a secondary effect.

The transition from highly generic to very specific ensembles in both cases is characterized by well defined phase transitions. These are visible in a number of network measures. Notably in both cases we have a rise of the largest eigenvalue of the adjacency matrix. This, however corresponds to the growth of the first dense region in the $R_S$-ensemble and to the emergence of an inter-regional hub in the $R_{WL}$-ensemble.

It is somewhat surprising that new things are still to learn on properties thought to characterize small-world networks. The fact that our perspective on relative canonical network ensembles could discover novel features is a promising sign for the study of properties of greater practical interest. In companion papers we are considering epidemic thresholds, and the vulnerability to cascading failures. More generally this method is of great interest wherever we want to understand and design topologies that fulfill certain functions, rather than describe empirical networks.

\subsection*{Code and Data availability}

All code and data used in this work will be made available at \url{https://doi.org/10.5281/zenodo.4462634}.

\appendix

\section{Relative Entropy}\label{Sec::Relative Entropy}

The minimization of the relative entropy has an information theoretic interpretation. Given a distribution $q$, the asymptotic probability to obtain a sample that looks like $p$ goes as the exponential of the negative relative entropy $D(p||q)$. This result of Chernoff \cite{chernoff1952measure} is known as Stein's Lemma (for a modern account phrased in terms of relative entropy see e.g. \cite{jaksic2018lectures} Theorem 4.12) and forms the mathematical basis for the interpretation of the relative entropy as a measure of distinguishability of probability distributions. Our ensembles thus have an information theoretic interpretation as being the ensembles that are hardest to distinguish from the generic ensemble $q$. In particular we do not presuppose that real network formation processes maximize entropy subject to some constraints, and do not interpret the resulting ensembles as modeling real networks that have the property $R$.

For completeness, we recall here the standard argument that the relative entropy, or the Kullback-Leibler divergence, is minimized by the exponential ensemble. We are looking for

\begin{align}
p^* &= \argmin_{\substack{p \\ \langle R \rangle = R^*}} D(p || q)\nonumber\\
    &= \argmin_{\substack{p \\ \langle R \rangle = R^*}} \sum_i p_i \ln\left(\frac{p_i}{q_i}\right)
\end{align}

First, note that this formula diverges to positive infinity if $p$ has support outside the support of $q$. We thus only consider $p$ whose support is contained in that of $q$. Then, by introducing Lagrange multipliers for the expectation value of $R$ as well as for the normalization condition on the distribution $p$ we can rewrite the constrained minimization above as a free minimization:

\begin{align}
    p^*(\beta_n, \beta_R) &= \argmin_{p} \sum_i p_i \ln\left(\frac{p_i}{q_i}\right) ~ + \nonumber\\
    &\phantom{=} + \beta_n \left(\sum_i p_i - 1\right) + \beta_R \left(\sum_i p_i R_i - R^*\right)\\
    R^* &= \sum_i R_i p_i^*(\beta_n, \beta_R)\nonumber\\
    1 &= \sum_i p_i^*(\beta_n, \beta_R) \nonumber
\end{align}

Now the variation in the direction $p_j$ produces the following condition:
\begin{align}
    0 &= \frac{\partial}{\partial p_j} \left [ \sum_i p_i (\ln (p_i) - \ln(q_i)) ~ + \right.\nonumber\\
    &\phantom{=} + \left. \beta_n \left(\sum_i p_i - 1\right) + \beta_R \left(\sum_i p_i R_i - R^*\right)\right] \nonumber \\
    &= \ln(p_j) - \ln(q_j) + 1 + \beta_n + \beta_R R_j
\end{align}

From which we can conclude

\begin{align}
    p^*_j &= \exp( \ln(q_j) - 1 - \beta_n - \beta_R R_j )\nonumber\\
    &= \frac 1 Z \, e^{ - \beta_R R_j } \, q_j
\end{align}

with $Z = e^{1 + \beta_n} = \sum_i e^{ - \beta_R R_i } \, q_i$ fixed by the condition $\sum_i p^*_i = 1$ and $\beta_R$ determined implicitly by the condition $R^* = \sum_i R_i p_i^*$.

Note that 
\begin{align}
    \frac{\partial}{\partial \beta_R}  \langle R \rangle &= - \frac 1 Z \sum_i R_i^2 e^{ - \beta_R R_i } \, q_i \nonumber\\
    &\phantom{=} \; - \frac{\partial Z}{\partial \beta_R} \frac 1 {Z^2} \sum_i R_i e^{ - \beta_R R_i } \, q_i \nonumber \\
    &= - \langle R^2 \rangle + \langle R \rangle^2\nonumber\\
    &= - \mathrm{Var}(R) \leq 0 \; .
\end{align}

Further, for $\beta_R = -\infty$ we have the distribution peaked completely on the global maxima: $\langle R \rangle = R_{\text{max}}$ and for $\beta_R = +\infty$ we have the minima instead $\langle R \rangle = R_{\text{min}}$. For $\beta_R = 0$ we have exactly the expectation value of $R$ in the generic background ensemble $q$.

\bibliography{bibliography}
\end{document}